\documentclass[10pt,preprintnumbers,twocolumn,nofootinbib]{revtex4}

\usepackage{epsfig}
\usepackage{amssymb}
\usepackage{amsmath}
\usepackage{mathrsfs}

\makeindex \textheight=9.5in \textwidth=6.5in \topmargin=-0.5in
\begin{document}
\title{Geometrically Engineering the Standard Model:\\Locally Unfolding Three Families out of $E_8$}
\author{Jacob L. Bourjaily}
\email{jbourjai@princeton.edu}
\affiliation{Joseph Henry Laboratories, Princeton University, Princeton, NJ 08544}
\date{3$^{\mathrm{rd}}$ April 2007}

\begin{abstract}
This paper extends and builds upon the results of \cite{Bourjaily:2007ab}, in which we described how to use the tools of geometrical engineering to deform geometrically-engineered grand unified models into ones with lower symmetry. This top-down unfolding has the advantage that the relative positions of singularities giving rise to the many `low energy' matter fields are related by only a few parameters which deform the geometry of the unified model. And because the relative positions of singularities are necessary to compute the superpotential, for example, this is a framework in which the arbitrariness of geometrically engineered models can be greatly reduced.

In \cite{Bourjaily:2007ab}, this picture was made concrete for the case of deforming the representations of an $SU_5$ model into their Standard Model content. In this paper we continue that discussion to show how a geometrically engineered $\mathbf{16}$ of $SO_{10}$ can be unfolded into the Standard Model, and how the three families of the Standard Model uniquely emerge from the unfolding of a single, isolated $E_8$ singularity.
\end{abstract}

\maketitle
\vspace{-0.0cm}\section{Introduction\label{intro}}\vspace{-0.0cm}
In \cite{Katz:1996xe}, Katz and Vafa showed how to geometrically engineer matter representations in terms of the local singularity structure of type IIa, M-theory, and F-theory compactifications. In that framework matter and gauge theory both have purely geometrical origins: $SU_n$, $SO_{2n}$ and $E_n$ gauge theories arise from the existence of co-dimension four singular curves of certain types in the compactification manifold \cite{Klemm:1995tj}; and massless matter representations arise from isolated points (in type IIa or M-theory) or curves (in F-theory) along the singular surface over which the type of singularity is enhanced by one rank.

Despite the extraordinary generality of this framework, it has not been widely used phenomenologically. This is largely because the description of the isolated enhancements of singularities giving rise to various matter representations is inherently local: although the geometry near any particular enhancement could be described concretely, the framework had nothing to say about numbers, types, and relative locations of different matter fields. This global data was either to be determined by duality to a concrete, global string theory model\footnote{Geometrically engineered models in M-theory are dual to intersecting brane models in type II, (see e.g. \cite{Acharya:2004qe}).}, or suggested via the {\it a posteriori} success of a given set of relative positions (as in e.g. \cite{Witten:2001bf,Friedmann:2002ty}).

\begin{figure*}[t]~\hspace{2.cm}~\includegraphics[scale=0.9]{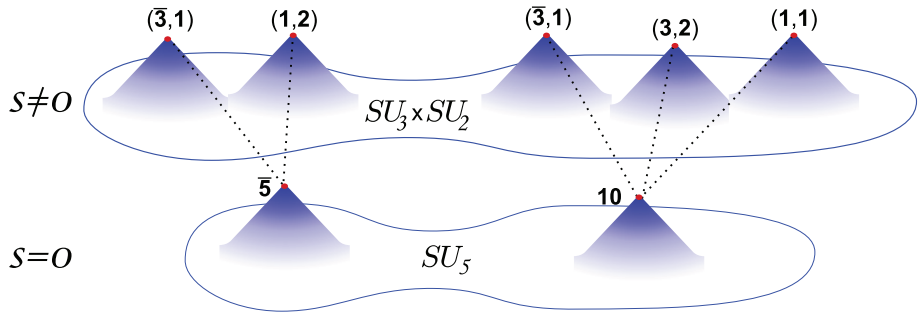}\caption{A cartoon of the geometric deformation of a $\overline{\mathbf{5}}$ and $\mathbf{10}$ of $SU_5$ into the Standard Model as described in \cite{Bourjaily:2007ab}. The surface over which the singularities are enhanced is coordinatized by a complex parameter $t$ and the geometry is deformed by changing the value of a complex parameter $s$. The relative locations of the `resolved' singularities are given in terms of $s$.}\label{sm_res_of_su5}\end{figure*}

Another way to relate the number and relative positions of (enhanced singularities giving rise to) matter fields was given in \cite{Bourjaily:2007ab}: in that paper, we described for example how a local description of the geometry giving rise to a massless $\overline{\mathbf{5}}$ of $SU_5$ could be smoothly deformed into a local description of a $(\overline{\mathbf{3}},\mathbf{1})$ and a $(\mathbf{1},\mathbf{2})$ of $SU_3\times SU_2$ which live at distinct points---related by a single deformation parameter. A cartoon of what was described in that paper is shown in Figure \ref{sm_res_of_su5}.

In this paper we describe pedagogically how to extend that idea to engineer analogies to $SO_{10}$ and $E_6\times SU_2$ grand unified models\footnote{As described in section \ref{e6xsu2}, the resolution \mbox{$E_8\to E_6\times SU_2$} naturally starts as a theory with three $\mathbf{27}$'s of $E_6$ related by an $SU_2$ family symmetry.}. Although in \cite{Bourjaily:2007ab} we were able to analyze explicit unfoldings of $SO_{10}$ and $SU_6$ singularities sufficiently well by sight, this will not be possible for our present examples. All of the examples in this paper involve the unfolding of isolated $E_n$ singularities; and although algebraic descriptions of these are known and classified \cite{Katz:1992ab}, it would be unnecessarily cumbersome and unenlightening to analyze them explicitly as we did in \cite{Bourjaily:2007ab}. Therefore, in \mbox{section \ref{resolvingEn}} we describe a much more powerful and elegant language in which to study these resolutions.

In section \ref{so10} we describe in detail how the unfolding of a $\mathbf{16}$ of $SO_{10}$ into the Standard Model is derived in the language of section \ref{resolvingEn}. This is achieved in two stages: in the first stage, we unfold the $\mathbf{16}$ into $\mathbf{10}\oplus\overline{\mathbf{5}}\oplus\mathbf{1}$ of $SU_5$; we then unfold the resulting $SU_5$ model into a single `family' of the Standard Model. At the end, all the relative positions of the singularities of the family are set by the non-zero values of two complex structure moduli, thereby greatly reducing the arbitrariness of their relative positions.

The next most obvious example would be a description of how a $\mathbf{27}$ of $E_6$ geometrically unfolds into the Standard Model. However, there are two reasons to leave this example to the reader: first, it is a most natural extension of the results of section \ref{so10}; secondly, it is a consequence of the $E_6\times SU_2$ grand unified model which we describe in section \ref{e6xsu2}. Although not given as an example in \cite{Katz:1996xe}, it is not hard to see\footnote{This is described in section \ref{resolvingEn}.} that a single isolated $E_8$ singularity at the intersection of a co-dimension four surfaces of types $E_6$ and $SU_2$ gives rise to matter in the representation $(\mathbf{27},\mathbf{2})\oplus(\mathbf{27},\mathbf{1})\oplus(\mathbf{1},\mathbf{2})$. It is easy to see how this would unfold into the matter content of three families---one coming from each of the $\mathbf{27}$'s. That three families emerge from $E_8$ is a general consequence of group theory and can be understood from the fact that $E_6\times SU_3$ is a maximal subgroup of $E_8$ into which the adjoint of $E_8$ partially branches into an $SU_3$ triplet of $\mathbf{27}$'s.

As in the preceding paper \cite{Bourjaily:2007ab}, this work is presented concretely in the language of Calabi-Yau compactifications of type IIa string theory, which can also be naturally extended to F-theory models. Here, we engineer the explicit local geometry of (non-compact) Calabi-Yau three-folds which are $K3$-fibrations over $\mathbb{C}^1$. If type IIa string theory is compactified on this three-fold, a four-dimensional $\mathscr{N}=2$ theory with various massless hypermultiplets will result. But if, for example, the $\mathbb{C}^1$ base of this three-fold were fibred as an $\mathscr{O}(-2)$ bundle over $\mathbb{CP}^1$, the resulting total space would be a Calabi-Yau four-fold\footnote{This is just one example of the ways in which these Calabi-Yau three-folds could be fibred over $\mathbb{CP}^1$ to result in a Calabi-Yau four-fold.} upon which F-theory would compactify to an $\mathscr{N}=1$ theory with chiral multiplets. However, because the manifold over which the singular $K3$'s are fibred in M-theory is a real, three-dimensional space, our fibrations over $\mathbb{C}^1$ do not have a direct application to M-theory.

It would of course be desirable to have a similar description of geometric unfolding explicitly in the language of $G_2$-manifolds so that this picture could be realized concretely in M-theory as well. This is particularly important in light of the recent advances in M-theory phenomenology (e.g. \cite{Acharya:2006ia,Acharya:2007rc}). By extension of the work of Berglund and Brandhuber in \cite{Berglund:2002hw}, such a generalization should be relatively straight-forward, but we will not attempt to do this here.
\vspace{-0.0cm}
\section{Resolving $E_n$-type Singularities\label{resolvingEn}}\vspace{-0.0cm}
Recall that a gauge theory in type IIa string theory can arise from compactification to six dimensions over a singular $K3$ surface (similar statements apply to M-theory and F-theory) \cite{Klemm:1995tj}. The complex structures of the singular compactification manifolds giving rise to $SU_n(\equiv A_{n-1})$, $SO_{2n}(\equiv D_n)$, and $E_n$ gauge theory are given in Table \ref{orbifolds}---where the surfaces are labelled conveniently by the name of the resulting gauge theory\footnote{It is of curious historical interest that the equations listed in Table \ref{orbifolds} were first identified by Fleix Klein in 1884 \cite{Klein:1884}. The reader may also be amused that the full resolutions of these surfaces were almost completely classified---up to a few computational errors---by Bramble in 1918 \cite{Bramble:1918}.}.

\begin{table}[b]
\begin{tabular}{lcr}
\toprule
Gauge group&{~}&Polynomial\\
\hline $SU_n$ ($\equiv A_{n-1}$) && $xy=z^n$\\
$SO_{2n}$ ($\equiv D_{n}$) && $x^2+y^2z=z^{n-1}$\\
$E_6$ && $x^2=y^3+z^4$\\
$E_7$ && $x^2+y^3=16yz^3$\\
$E_8$ && $x^2+y^3=z^5$\\\botrule
\end{tabular}\caption{\label{orbifolds}Hypersurfaces in $\mathbb{C}^3$ giving rise to the desired orbifold singularities.}\end{table}

We can generalize this discussion by considering a complex, one-dimensional space $B$ over which a smooth family of singular $K3$ surfaces are fibred. If almost everywhere over $B$ the $K3$-fibres have singularities of a single type, then compactification of type IIa string theory over the total space will give rise to gauge theory in four-dimensions of the type corresponding to the typical fibre. Massless charged matter will arise if over isolated points in $B$ the type of fibre is enhanced by one rank. The geometry about a single such isolated point where the singularity is enhanced was described in detail by Katz and Vafa in \cite{Katz:1996xe}.

The representation of matter living at these `more-singular' points was also given in \cite{Katz:1996xe}: suppose that $G\supset H\times U_1$ and that the rank of $H$ is one less than $G$; then, if there is an isolated $G$-type singularity over a surface of $H$-type singularities, the resulting massless representation is given by those parts of the decomposition of the adjoint of $G$ into $H\times U_1$ which are charged under the $U_1$.

Because the question of how to (smoothly) deform the surfaces of Table \ref{orbifolds} into ones of lower rank has intrinsic mathematical interest, it is not too surprising that all possible two-dimensional deformations have been classified. Our discussion below will make use of the notation and results presented in \cite{Katz:1992ab}.

In our present work, we are interested in deformations of $E_n$ singularities into ones of lower rank. Unlike $SU_n$ singularities, the resolutions of which are easy enough to read off by sight, the algebraic complexity of $E_n$ singularities is formidable. To appreciate what is meant by this, consider the resolution of $E_7$. From Table \ref{orbifolds} we know that an $E_7$ singularity is locally isomorphic to the surface $x^2+y^3=16yz^3$ in $\mathbb{C}^3$. Its full resolution in terms of the seven deformation parameters $\vec{\!{~}\!t}=(t_1,t_2,\ldots,t_7)$ is given by \begin{widetext} \begin{equation}-x^2-y^3+16yz^3+\epsilon_2(\vec{\!{~}\!t})y^2z+\epsilon_6(\vec{\!{~}\!t})y^2+\epsilon_8(\vec{\!{~}\!t})yz+\epsilon_{10}(\vec{\!{~}\!t})z^2+\epsilon_{12}(\vec{\!{~}\!t})y+\epsilon_{14}(\vec{\!{~}\!t})z+\epsilon_{18}(\vec{\!{~}\!t})=0,\label{e7res}\end{equation} \end{widetext} where the $\epsilon_n(\vec{\!{~}\!t})$ are $n^{\mathrm{th}}$ order symmetric polynomials in the components of $\vec{\!{~}\!t}$ which are tabulated over several pages of the appendix of \cite{Katz:1992ab}.

A na\"{i}ve way to determine the type of singularity found by resolving $E_7$ ``in the direction $\vec{\!{~}\!t}$'' would be to expand equation (\ref{e7res}) completely using the explicit functions $\epsilon_n(\vec{\!{~}\!t})$, find each of its singular points, and expand locally about each until an isomorphism with a singularity of lower rank in Table \ref{orbifolds} was clear. This is the way, for example, that \cite{Katz:1996xe} demonstrated that the resolution of $E_7$ in the direction $(0,0,0,0,0,t,0)$ gives rise to $E_6$ for $t\neq0$. All of the results in this paper could be verified in this way. Luckily, however, Katz and Morrison described a much more powerful and direct way to analyze the deformations of $E_n$ singularities \cite{Katz:1992ab}.

We would like a pragmatic answer to the following question: {\it what is the type of fibre found by resolving an $E_n$ singularity in the direction $\vec{\!{~}\!t}$?} That there is an easy answer to this question makes our work much simpler. Although an adequate treatment would take us well beyond the scope of our present discussion, the answer given in \cite{Katz:1992ab} is at least very easy to make use of\footnote{Of course, this answer does depend on the parameterization used. As stated before, we are working with the conventions of \cite{Katz:1992ab}.}: {\it for each of the equations in Table \ref{roots} satisfied by the components of $\vec{\!{~}\!t}$, the singularity has the corresponding root.} Given the list of roots, it is then a straight-forward exercise to construct the Dynkin diagram corresponding to the singularity\footnote{Misusing the notation of \cite{Katz:1992ab} in a way applicable only to $SU_n$, $SO_{2n}$ and $E_n$, one can think of the vectors $e_i$ as an orthonormal basis in Minkowski space which is equipped with a mostly-plus metric. Then roots are vectors in this space of norm $+2$. Each (positive) root gives rise to a node in the resulting Dynkin diagram, and two nodes are connected by a line if their inner product is $-1$ and disconnected if they are orthogonal.}.

In an admittedly bad notation, we consider each of the $n$ deformation parameters $t_i(t)$ to be functions of $t$, the local coordinate on the base space $B$. A (non-Abelian) gauge theory will be present if there are roots implied by Table \ref{roots} which are preserved for generic values of $t$. And charged massless matter will exist if at isolated points $\{t_*\}$ an additional root is added---or, in terms of Dynkin diagrams, if an additional node is added. At each isolated point we can therefore identify the resolution $G\to H$ and thereby determine the resulting representation.

\begin{table}[b]
\begin{tabular}{ccc}
\toprule
Equation&&Root\\
\hline $t_i-t_j=0$&$\implies$&$e_i-e_j$\\
$t_i+t_j+t_k=0$&$\implies$& $e_0-e_i-e_j-e_k$\\
$\sum_{j=1}^6t_{i_j}=0$ &$\implies$& $2e_0-\sum_{j=1}^{6}e_{i_j}$\\
{~}\hspace{0.25cm}$2t_{i_1}+\sum_{j=2}^7t_{i_j}=0$&$\implies$& $3e_0-2e_{i_1}-\sum_{j=2}^7e_{i_j}$\\
\botrule
\end{tabular}\caption{\label{roots}The roots of the singularity resulting from the resolution of $E_n$ in the direction $\vec{\!{~}\!t}$. This is a reproduction of Table 4 of Ref. \cite{Katz:1992ab}.}\end{table}

\newpage
\section{Geometric Analogue of $SO_{10}$ Grand Unification\label{so10}}
\subsection{The Description of a $\mathbf{16}$ of $SO_{10}$}\vspace{-0.35cm}
A necessary starting point to describe the unfolding of a $\mathbf{16}$ of $SO_{10}$ into the Standard Model is a description of the initial geometry as was done in \cite{Katz:1996xe}. We will briefly review that construction in the language described above before we unfold it, first into an $SU_5$ model, and later all the way into $SU_3\times SU_2$.

Let $t$ be a local complex coordinate on the space $B$ over which is fibred the resolution of $E_6$ parameterized by $\vec{\!{~}\!t}=(t,t,t,t,t,-2t)$. To be clear, for each value of $t$, the vector $\vec{\!{~}\!t}$ describes an explicit surface in $\mathbb{C}^3$ given in reference \cite{Katz:1992ab} analogous to that of equation (\ref{e7res}) above.

Considering the rules of Table \ref{roots}, we see that for an arbitrary value of $t\neq0$ the root lattice of the fibre is \begin{equation}\left\{\begin{array}{c}(e_0\!-\!e_1\!-\!e_2\!-\!e_6)\hspace{1.72cm}{~}\\(e_1-e_2)\quad(e_2-e_3)\quad(e_3-e_4)\quad(e_4-e_5)\end{array}\right\},\end{equation}
where we have displayed the roots suggestively so as to reproduce the $SO_{10}$ Dynkin diagram. At $t=0$, however, $E_6$ is restored. So we have an isolated $E_6$ fibre over the point $t=0$, while for any $t\neq0$ the fibre is $SO_{10}$. This gives rise to $SO_{10}$ gauge theory with a single massless $\mathbf{16}$ located at the origin in the $t$-plane.

\subsection{Unfolding the $\mathbf{16}$ of $SO_{10}$ into $SU_5$}
We would like to unfold the manifold described above into one with $SU_5$ gauge theory. It is not hard to guess in what `directions' $\vec{\!{~}\!t}$ we may deform the the geometry so that the fibre over a generic point is $SU_5$. Let $a$ denote a parameter independent of $t$ which adjusts the whole geometry over the region which is coordinatized by $t$. Then let the fibre over $t$ be given by the resolution of $E_6$ in the direction $(t,t,t,t,t+a,-2t-a)$. Obviously when $a=0$ the situation is the same as above and results in a single massless $\mathbf{16}$ of $SO_{10}$. However, when \mbox{$a\neq0$} the situation is different: for generic values of $t$ it is easy to see that the simple roots are \begin{equation}\Big\{(e_0-e_1-e_5-e_6)\quad(e_1-e_2)\quad(e_2-e_3)\quad(e_3-e_4)\Big\},\end{equation} which means that the generic fibre over $t$ is just $SU_5$---and so the resulting gauge theory is $SU_5$.

To find what matter representations exist, we must determine over which locations $t$ the rank of the fibre is enhanced. This means we are seeking special values of $t$ (determined by $a$) at which an additional equation in Table \ref{roots} is satisfied. For each of these points, we can draw the resulting Dynkin diagram to determine the fibre over that point, thereby determining the representation which arises there.

\begin{table}[t]\caption{\label{so10unfolding1}The locations on the complex $t$-plane over which the singularity of the fibre is enhanced, and the representations of $SU_5\times U_1$ that result.}
\begin{tabular}{rcc}\toprule Location&Fibre&$\begin{array}{c}\text{Representation}\\\text{of}~SU_5\times U_1\end{array}$\\\hline
$3t+2a=0$&$\quad SU_5\times SU_2$&$\mathbf{1}_{-5}$\\
$3t+a=0$&$SO_{10}$&$\mathbf{10}_{-1}$\\
$t=0$&$SU_6$&$\overline{\mathbf{5}}_{3}$\\\botrule
\end{tabular}
\end{table}

It is not hard to exhaustively find all these `more singular' points. They are give in \mbox{Table \ref{so10unfolding1}}. Notice that we have included the $U_1$-charge assignments that result; these are normalized as in the appendix of \cite{Slansky:1981yr}.

\subsection{Unfolding a $\mathbf{16}$ of $SO_{10}$ into the Standard Model}

To complete our task and unfold the $\mathbf{16}$ of $SO_{10}$ all the way to the Standard Model, we must deform the fibres by another `global' parameter, which we will denote $b$. It is not hard to guess a direction over which the generic fibre will be $SU_3\times SU_2$: try for example $(t,t,t,t+b,t+a,-2t-a-b)$. Again, we notice that for a general location $t$ and generic fixed values $a,b\neq0$, the singularity has the root structure \begin{equation}\Big\{(e_1-e_2)\quad(e_2-e_3)\Big\}\otimes\Big\{(e_0-e_4-e_5-e_6)\Big\},\end{equation}
which is visibly $SU_3\times SU_2$.

\begin{figure*}[t]\includegraphics[scale=1]{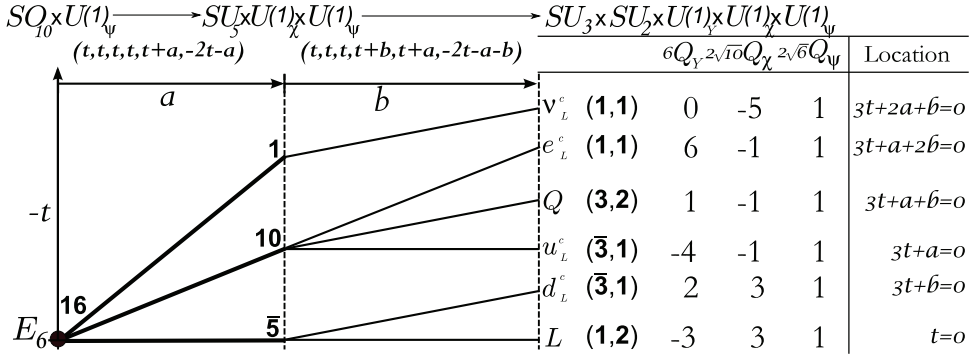}\caption{An illustration of the resolution of a geometrically engineered $\mathbf{16}$ of $SO_{10}$ into the Standard Model as a function of two complex structure moduli $a$ and $b$ as described in section \ref{so10}. The coordinate along the base space is $t$ and runs vertically in the diagram. When $a=b=0$, along the left-hand side, there is just one isolated $E_6$ singularity at $t=0$. When $b=0$ but $a$ is allowed to vary, this single singularity splits into three, and any $a=$constant slice will have three isolated singularities in the complex $t$-plane as shown above. Moving rightward in the diagram, at the dashed line $a$ is held fixed and $b$ is allowed to grow, causing the three enhancements of $SU_5$ to break apart into six total isolated singularities over $SU_3\times SU_2$, which is shown on the right-hand-side. Also shown are the (appropriately normalized) $U_1$ charges of fields obtained via this multiple unfolding.}\label{so10tosmdiagram}\end{figure*}

Like above, it is a straight-forward exercise to determine all the locations over which the singularity is enhanced, and the resulting representation which arises. These points including their resulting representations (with $U_1$-charges as normalized in \cite{Slansky:1981yr}) are listed in Table \ref{so10unfolding2}. The entire unfolding is reproduced graphically in \mbox{Figure \ref{so10tosmdiagram}}.

\begin{table}[b]\squeezetable
\begin{tabular}{rccc}\toprule Location&Fibre&$\substack{\text{Representation}\\\text{of}~SU_3\times SU_2\times U_1}$&Name\\\hline
$3t+2a+b=0$&$SU_3\times SU_2\times SU_2$&$(\mathbf{1},\mathbf{1})_{0}$&$\nu^c_L$\\
$3t+a+2b=0$&$SU_3\times SU_2\times SU_2$&$(\mathbf{1},\mathbf{1})_{6}$&$e^c_L$\\
$3t+a+b=0$&$SU_5$&$(\mathbf{3},\mathbf{2})_{1}$&$Q$\\
$3t+a=0$&$ SU_4\times SU_2$&$(\overline{\mathbf{3}},\mathbf{1})_{-4}$&$u^c_L$\\
$3t+b=0$&$SU_4\times SU_2$&$(\overline{\mathbf{3}},\mathbf{1})_{2}$&$d^c_L$\\
$t=0$&$SU_3\times SU_3$&$(\mathbf{1},\mathbf{2})_{-3}$&$L$\\
\botrule
\end{tabular}\caption{\label{so10unfolding2}The locations on the complex $t$-plane over which the singularity of the fibre is enhanced and the representations of $SU_3\times SU_2\times U_1$ that result.}
\end{table}

\newpage
\section{Geometric Analogue of $E_6\times SU_2$ Grand Unification\label{e6xsu2}}
After having completed the unfolding of a $\mathbf{16}$ of $SO_{10}$ into the Standard Model, it is natural to ask if this idea can be extended to relate all the singularities of the Standard Model as perhaps the unfolding of a single isolated singularity of higher-rank. The answer is in fact yes---and there is a sense in which precisely three families arise if the notion of `geometric unification' is saturated.

Because a $\mathbf{16}$ of $SO_{10}$ arises from the resolution $E_6\to SO_{10}$, it can only be unfolded out an exceptional singularity. Clearly the highest level of unification one can achieve along this line would be to start with a resolution $E_8\to H$ where $H$ is a rank-seven subgroup of $E_8$ which contains $SO_{10}$. The possible `top-level' gauge groups are then $E_7$, $E_6\times SU_2$, and $SO_{10}\times SU_3$. We choose to study \mbox{$E_8\to E_6\times SU_2$} as our example because it will naturally include a description of the unfolding of $\mathbf{27}$ of $E_6$ into the Standard Model, which is interesting in its own right, and because it follows quite directly from our work in section \ref{so10}.

The initial geometry which we will deform into the Standard Model is given as follows. Let $t$ be a complex coordinate on the base space $B$ over which is fibred the resolution $(t,t,0,0,0,0,0,0)$ of $E_8$. Clearly, when $t=0$ we recover $E_8$; when $t\neq0$ we see that the roots of the fibre are \begin{widetext}\begin{equation}\left\{\begin{array}{c}(e_0\!-\!e_3\!-\!e_4\!-\!e_5)\\(e_3-e_4)\quad(e_4-e_5)\quad(e_5-e_6)\quad(e_6-e_7)\quad(e_7-e_8)\end{array}\right\}\otimes\Big\{(e_1-e_2)\Big\},\label{e6xsu2dynkin}\end{equation}\end{widetext}
which is visibly $E_6\times SU_2$. Following the general rule to determine the representation resulting from a given resolution \cite{Katz:1996xe}, we find that at $t=0$ lives massless matter charged in the $(\mathbf{27},\mathbf{2})_{1}\oplus(\mathbf{27},\mathbf{1})_{-2}\oplus(\mathbf{1},\mathbf{2})_{3}$ representation of $E_6\times SU_2\times U_{1_{\varphi}}$.

\begin{figure*}[!]\includegraphics[scale=0.75]{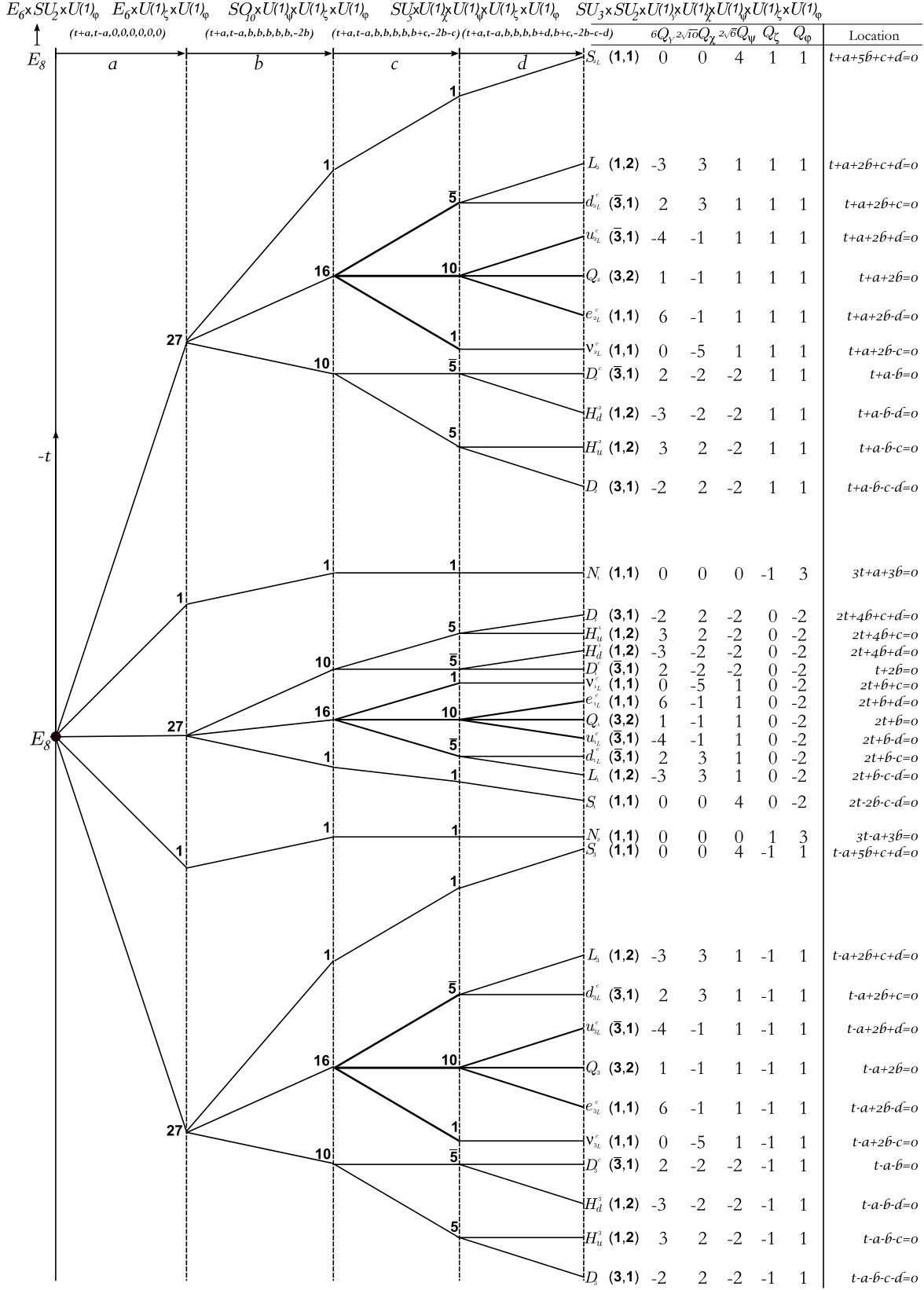}\caption{An illustration of the resolution of a single isolated $E_8$ into the Standard Model in terms of four deformation parameters $a,b,c,d$. Along the left hand side, for $a=b=c=d=0$, the generic $E_6\times SU_2$ fibre is enhanced to $E_8$ at $t=0$. Moving from left to right, $a,b,c,d$ are sequentially allowed to grow to some non-zero value---and between dashed lines all but one of the moduli are held fixed. Solid lines indicate the locations of enhanced singularities relative to the plane for as functions of $a,b,c,d$. The complete list of isolated singularities, their locations, and charge assignments are given on the right hand side of the diagram.}
\label{fullres}\end{figure*}

To avoid pedantic redundancy, in Figure \ref{fullres} we have summarized in great detail the entire unfolding into $SU_3\times SU_2\times U_{1_Y}\times U_{1_{\chi}}\times U_{1_{\Psi}}\times U_{1_{\zeta}}\times U_{1_{\varphi}}$. An outline of the steps involved in deriving this unfolding is given presently.

First, the unfolding of the $E_6\times SU_2$ gauge theory into $E_6$ gauge theory is obtained by defining the fibre over $t$ to be given by the resolution of $E_8$ in the direction $(t+a,t-a,0,0,0,0,0,0)$ for some \mbox{$a\neq0$}. This clearly kills the $SU_2$ node of the fibre in \mbox{equation (\ref{e6xsu2dynkin})}. There are five locations at which the singularity is enhanced by one rank, giving rise to three $\mathbf{27}$'s and two singlets as shown in the left-most section of \mbox{Figure \ref{fullres}}.

\newpage The rest of the unfolding is a natural application of the work in section \ref{so10}. Let us now set the fibre over $t$ to be given by the resolution \mbox{$(t+a,t-a,b,b,b,b,b,-2b)$} of $E_8$ for arbitrary complex deformation parameters $a,b\neq0$. From section \ref{so10} we see immediately that the generic fibre is $SO_{10}$. A thorough scanning for possible solutions to equations in Table \ref{roots} shows that there are $11$ isolated points on the complex $t$-plane over which the singularity is enhanced. These correspond to the `breaking' of each $\mathbf{27}$ of $E_6$ into $\mathbf{16}\oplus\mathbf{10}\oplus\mathbf{1}$ of $SO_{10}$, while the singlets remain singlets. This is seen in the second vertical strip (from the left) in Figure \ref{fullres}.

Again, following our discussion above, it is easy to guess possibilities for the next two resolution directions. First, we set the fibre over $t$ to be given by $(t+a,t-a,b,b,b,b,b+c,-2b-c)$ which will result $SU_5$ gauge theory with matter content corresponding to the `canonical' decomposition of three $\mathbf{27}$'s of $E_6$ with two singlets. And finally, the full resolution of the $E_6\times SU_2$ grand unified model into $SU_3\times SU_2$ can be given by letting the fibre over $t$ be given by the $(t+a,t-a,t,t,t,t+d,t+c,-2b-c-d)$ resolution of $E_8$ for (generic) arbitrary fixed complex structure moduli $a,b,c,d\neq0$.
\vspace{-0.5cm}\section{Implications}\vspace{-0.5cm}
Let us clarify what we have done. For a given set of fixed, nonzero complex structure moduli, the resolution given above describes the explicit, local geometry of a non-compact Calabi-Yau three-fold, which is a $K3$-fibration over $\mathbb{C}^1$. If type IIa string theory is compactified on this three-fold, the resulting four-dimensional theory will have $SU_3\times SU_2$ gauge theory with hypermultiplets at isolated points as given in Figure \ref{fullres} which reproduce the spectrum of three families of the Standard Model with an extended Higgs sector and some exotics. Alternatively, if one takes this (non-compact) Calabi-Yau three-fold and fibres it over $\mathbb{CP}^1$ as described in section \ref{intro} so that the total space is Calabi-Yau, then F-theory on this space will give rise to $\mathscr{N}=1$ supersymmetry with $SU_3\times SU_2$ gauge theory and chiral multiplets in the representations given in Figure \ref{fullres}. And although it does not follow directly from our construction above, considering the close similarities between two- and three-dimensional resolutions of the singular $K3$ surfaces we have every reason to suspect an analogous geometry can be engineered for M-theory in terms of hyper-K\"{a}hler quotients by extension of the results in \cite{Atiyah:2001qf,Acharya:2001gy,Berglund:2002hw}. We are currently working on building this geometry in M-theory, and we expect to report on this work soon.

Given these four complex structure moduli, all the relative positions of the 35 disparate singularities giving rise to all three families of the (extended) Standard Model are then known\footnote{A subtlety, however, is that because our language has been explicitly that of $\mathscr{N}=2$ theory from type IIa, we are unable to distinguish the $\mathbf{5}$ from the $\overline{\mathbf{5}}$ in the splitting of the $\mathbf{10}$'s of $SO_{10}$. In Figure \ref{fullres}, a consistent choice was made---and although we do not justify this claim here, it is the choice that will be correct for the M-theory generalization of this work.}. Beyond the usual three families of the Standard Model, the manifold also gives rise to two Higgs doublets for each family, six Higgs colour triplets, three right-handed neutrinos and five other Standard Model singlets. We should point out that this matter content (and their $U_1$-charge assignments) is a consequence of group theory and algebraic geometry alone---it is simply what is found when unfolding $E_8$ all the way to the Standard Model.

And given the relative positions and local geometry of the singularities together with the $U_1$-structure, one can in principle compute the full superpotential coming from instantons wrapping different singularities. Because these are fixed by the values of the complex structure moduli, there is a (complex) four-dimensional landscape\footnote{That it is continuous is a consequence of the fact that we are engineering non-compact Calabi-Yaus. If one matched this local geometry to a compact global structure, the landscape would of course be discrete.} of different, explicit $SU_3\times SU_2$ embeddings at the compactification scale. Although this large landscape may appear to have too much freedom, we remind the reader that in the traditional understanding of geometrical engineering there would be hundreds of parameters describing the (independent) relative locations of each of the isolated singularities.

There are a few things to notice about the form of the superpotential that will emerge. First, because of the $U_{1}$-charge assignments, each term in the superpotential must combine exactly one term arising from each of the $\mathbf{27}$'s. This greatly limits the form of the superpotential. And in particular, it implies that neither mass nor flavour eigenstates will arise from any single $\mathbf{27}$---that is, the `families' in the colloquial sense are necessarily linear combinations of fields resulting from different $\mathbf{27}$'s.

Also notice that in general the terms in the superpotential will be proportional to $e^{-\int\!\!d\mathrm{Vol}}$ where $d\mathrm{Vol}$ is the volume form of some cycle wrapping singularities (the details of which depends on whether we are talking about type IIa, M-theory, or F-theory realizations), and are in principle calculable in terms of the deformation moldui. And because these coefficients are exponentially related to the volumes of cycles, we expect the high-scale Lagrangian will be generically hierarchical. This structure could be important for solving problems in phenomenology---for example the $\mu$ problem in the Higgs potential, the Higgs doublet-triplet splitting problem, or avoiding proton decay.

We are in the process of studying the phenomenology of models on this landscape. At first glance, the $U_1$-structure combined with high-scale hierarchies could possibly be complex enough to be able to avoid some of the typical problems of $E_6$-like grand unified models. We should point out that if there were no high-scale hierarchies, however, then the allowed terms in the superpotential would generically give rise to low-energy lepton and baryon number violation, similar to any `generic' $E_6$ model---i.e. one which includes all types of terms allowed by the $E_6$-mandated $U_1$-structure \cite{Hewett:1988xc}. We could always impose additional symmetries and add fields by hand to solve these problems, but this would not be very compelling. However, if viable models already exist in the landscape which do not require additional fields or symmetries, these would be compelling even if we do not yet understand how they are selected.

One of the most important phenomenological questions about these models is the fate of the additional $U_1$ symmetries. Although we suspect that one can determine which of the $U_1$ symmetries are dynamical below the compactification scale by studying the normalizability of their corresponding vector multiplets, we do not presently have have a complete understanding of this situation. Of course, if any additional $U_1$'s survive to low energy they could have very interesting---or damning---phenomenological consequences.

\vspace{-0.5cm}\section{Discussion}\vspace{-0.35cm}
An important point to bear in mind when considering geometrically engineered models is that there generically exist\footnote{There could be global obstructions which prevent such a deformation from taking place. But these are invisible to the non-compact, local constructions considered here.} moduli which can deform the geometry into one which gives rise to a theory with less gauge symmetry. For example, if you are given a geometrically-engineered $SO_{10}$ grand unified model, then our results show explicitly that the model can be locally deformed into an $SU_5$ model, and this can be deformed further into the Standard Model; the original $SO_{10}$ theory is seen to be a single point in a (complex) two-dimensional landscape of $SU_3\times SU_2$ theories. And because larger symmetries always lie in lower dimensional surfaces of moduli space, it is very relevant to ask what physics prevents this unfolding from taking place. Indeed, this question applies to the Standard Model as well---our analysis could easily go further to unfold away $SU_3\times SU_2$. We are not presently able to answer why this does not happen\footnote{Although, perhaps the unfolding of $SU_2$ may provide an alternative to tuning in the usual Higgs sector \cite{Langacker:email}. It would be interesting to understand in greater detail the relationship between unfolding and the Higgs mechanism.}; although this observation suggests that perhaps theories with less symmetry, like $SU_3\times SU_2$, could be much more natural than grand unified theories.

More generally, it is not presently understood what physics controls the values of the geometric moduli which deform the manifold---the parameters which deform the $E_8\to E_6\times SU_2$ complex structure, for example. We do not yet have a general mechanism which would fix these parameters; we simply observe that any non-zero values of the moduli will give rise to a geometrically engineered manifold with $SU_3\times SU_2$ gauge theory `peppered' with all the necessary singularities of the three families of the Standard Model together with the usual $E_6$-like exotics. And importantly, for any point in the complex four-dimensional `landscape,' the relative locations of all the relevant singularities are known---and hence in principle so is the superpotential.

This relationship between moduli-fixing and gauge symmetry breaking could be a novel feature of geometrically-unfolded models. It may allow one to apply the results in \cite{Acharya:2006ia}, for example, to single out theories on the landscape. However, a prerequisite to this type of analysis would be an identification of which moduli should be identified with the ones which deform the geometry as described here.

Although the motivation in this paper and in \cite{Bourjaily:2007ab} appears to be a top-down realization of grand unification, there is a sense in which we are really engineering from the bottom-up. Specifically, because the local geometry we have described is non-compact, the resulting theory is decoupled from quantum gravity, and the parameters along the landscape of deformations are continuous. This is not unlike the situation in \cite{Verlinde:2005jr}. But what we lose in global constraints we perhaps gain by concrete local structure. Not only do we have a framework which naturally predicts three families with a rather detailed phenomenological structure, but we have done so in a way that preserves all the information about the local geometry. And because this framework realizes the `physics from pure geometry' paradigm in a potentially powerful way, it could prove important to concrete phenomenological constructions in M-theory, for example.

Of course we envision these local geometries to be embedded within compact Calabi-Yau manifolds. It is an assumption of the framework that the precise global topology of the compactification manifold can be ignored at least as a good first approximation. One may ask the extent to which these constructions can be glued into compact manifolds. Concretely: {\it under what circumstances can a non-compact Calabi-Yau three-fold which is a fibration of $K3$ surfaces with asymptotically uniform ADE-type singularities be compactified}? This is an important question for mathematicians, the answer to which would likely lead to important physical insight---e.g. quantization of the moduli space of deformations.

A possible objection to this framework is that our constructions appear to depend on several seemingly arbitrary choices (the specific chain from $E_8$ to the Standard Model, which roots were eliminated at each step, etc.). However, it is likely that the particle content, for example, which results is completely independent of these choices. Furthermore, we suspect that different realizations of the unfolding merely result in different parameterizations of the landscape, and do not reflect true additional arbitrariness. But this is still an area that deserves attention.

Lastly, because in this picture the Standard Model is seen to unfold at the compactification scale, one may ask what has become of gauge coupling unification. Because the gauge coupling constants are functions of the volumes of their corresponding co-dimension four singular surfaces\footnote{Of course, this can only be discussed concretely when the compact manifold is known.} which depend on the deformation moduli, the traditional meaning of grand unification is more subtle here---as is typical in string phenomenology. For example, although we chose to unfold the Standard Model sequentially as a series of less unified models, there is no reason to suspect that that order has any physical importance. Surely, if as we parameterized the unfolding in section \ref{e6xsu2}, setting $d\to0$ (or $c\to0$) would result in an $SU_5$ grand unified theory; but setting $a\to0$ instead would result in a restoration of family symmetry. The four complex structure moduli tune different types of unification separately---and should simultaneously be at play in the question of gauge coupling unification.

It is interesting to note, however, that if one were to simultaneously scale the values of all the moduli to be very small, the spectrum would be more and more unified: the relative distances between singularities shrink, unifying the coefficients in the superpotential; and the volumes of the co-dimension four singularities (if realized in a compact manifold) would approach one another, resulting in a unification of their gauge couplings. What this may mean phenomenologically remains to be understood.

In this paper we have described a local, purely geometric framework in which gauge symmetry `breaking' can be re-cast as a problem of moduli fixing---and in which the same moduli which describe this geometric `unfolding' also determine the physics of massless matter. And although we still do not understand the mechanisms by which these moduli are fixed, the landscape of possibilities is already enormously reduced: what would have been the hundreds of parameters describing the relative positions on the compactification manifold of the Standard Model's three families worth of matter fields, we specify them all in terms of only four complex structure moduli which describe the unfolding of an isolated $E_8$ singularity. And the fact that three families emerges is group-theoretic and not added by hand.

\vspace{-0.5cm}\section{Acknowledgements}
\vspace{-0.5cm}
It is a pleasure to thank helpful discussions with and insightful comments of Herman Verlinde, Sergei Gukov, Gordy Kane, Paul Langacker, Edward Witten, Cumrun Vafa, Brent Nelson, Malcolm Perry, Dmitry Malyshev, Matthew Buican, Piyush Kumar, and Konstantin Bobkov.

This work was funded in part by a Graduate Research Fellowship from the National Science Foundation.


\begin{thebibliography}{10}
\bibitem{Langacker:email}
 This was pointed out by Paul Langacker in a private correspondance.
\end{thebibliography}

\end{document}